\documentclass[ aip, amsmath, amssymb,
 reprint,%
]{revtex4-1}

\usepackage{graphicx}
\usepackage{dcolumn}
\usepackage{bm}

\usepackage[utf8]{inputenc}
\usepackage[T1]{fontenc}
\usepackage{mathptmx}
\usepackage{etoolbox}
\usepackage{color}

\begin{document}

\title{Sensitivity enhancement of an anomalous Hall effect magnetic sensor by means of second-order magnetic anisotropy}

\author{H. Arai}
\author{H. Imamura}%
\affiliation{
National Institute of Advanced Industrial Science and Technology (AIST), 1-1-1 Umezono, Tsukuba 305-8568, Japan
}
\email{Authors to whom correspondence should be addressed: [Hiroko Arai, arai-h@aist.go.jp; Hiroshi Imamura, h-imamura@aist.go.jp] }

\date{\today}

\begin{abstract}
The sensing performance of anomalous Hall effect (AHE) magnetic sensors is investigated 
in terms of their sensitivity, 
the power spectrum of their voltage noise, 
and their detectivity. 
Special attention is paid to the effect of the second-order anisotropy constant, 
$K_{2}$, 
on the sensing performance. 
It is found that 
the sensitivity is strongly enhanced 
by tuning the value of $K_{2}$ 
close to the boundary 
between the in-plane magnetized state 
and the conically magnetized state. 
It is also found that 
the detectivity is almost independent of $K_{2}$ 
as long as the film is in-plane magnetized. 
These results provide fundamental insights 
into the design of high-performance AHE sensors.
\end{abstract}

\maketitle

The anomalous Hall effect (AHE) has attracted much attention 
in the fields of physics 
\cite{smit_r1_06_01_the_1955,smit_r1_06_01_the_1958,berger_r1_06_01_side-jump_1970,onoda_r1_06_01_topological_2002,jungwirth_r1_06_01_anomalous_2002,xu_r1_06_05_effect_2024} 
and device applications.\cite{diao_r2_06_02_magnetic_2010,lu_r2_06_01_ultrasensitive_2012,zhu_r2_06_01_giant_2014,peng_r2_06_01_the_2019,zhang_r2_06_02_low-frequency_2019,wang_r2_06_01_anomalous_2020,zhang_r2_06_02_noise_2020,ramesh_r2_06_06_biological_2022,shiogai_r2_06_02_improvement_2022,nakatani_r2_06_02_perspective_2024}
One of the most promising device applications is an AHE magnetic sensor 
that can detect a very small magnetic field 
through the out-of-plane (OP) component of the magnetization 
in a magnetic thin film. 
Due to the detection principle, 
AHE sensors require magnetic thin films 
with an in-plane (IP) magnetized state 
or a magnetic state 
with the magnetic moment 
slightly tilted from the IP direction.
Much effort has been devoted to finding materials 
with large anomalous Hall resistivity
to enhance the sensitivity of AHE magnetic sensors.
Previous studies have investigated materials 
with large spin-orbital interactions, 
such as ferromagnetic alloys, 
\cite{zhang_r2_06_02_low-frequency_2019,shiogai_r2_06_02_improvement_2022} 
spintronics materials, 
\cite{zhu_r2_06_01_giant_2014,zhang_r2_06_02_noise_2020,ramesh_r2_06_06_biological_2022,xu_r1_06_05_effect_2024, wang_spintronic_2023,liu_domain_2024} 
and topological materials.
\cite{nakatani_r2_06_02_perspective_2024}

In this work, 
we propose another method of enhancing the sensitivity of an AHE magnetic sensor. 
The sensitivity of AHE magnetic sensors can be improved 
by improving the response of the magnetization 
to the external magnetic field to be detected, 
i.e., 
the signal field. 
The response of the magnetization of an AHE magnetic sensor is determined 
by the competition 
between the torque from the signal field 
and the torque derived from magnetic anisotropy, 
such as shape and crystalline anisotropy.
Some magnetic films have a large second-order uniaxial anisotropy constant, 
$K_{2}$, 
in the OP direction, 
resulting in the conically magnetized state, 
called the cone state, 
as an equilibrium state. 
Su et al. reported on the potential of AHE sensors with cone states 
for three-dimensional magnetic field detection. 
\cite{su_r3_06_02_easycone_2023}
It is known that, 
even in the case of an IP magnetized state, 
the magnetization response to the external field can be modified 
using $K_{2}$. 
The impact of $K_{2}$ on switching properties has been examined in materials 
utilized in magnetic recording for storage 
\cite{kitakami_energy_2003,shimatsu_k_2005} 
and memory 
\cite{matsumoto_efficiency_2017,matsumoto_voltage-induced_2018} 
applications. 
$K_{2}$ represents a pivotal parameter 
for regulating magnetic properties in a systematic manner.

In this study, 
using the macrospin model,
we analyze the effect of $K_{2}$ 
on the sensing performance of AHE magnetic sensors.
The sensitivity is obtained 
by calculating the magnetic field dependence 
of the stable direction of magnetization. 
The voltage noise power spectrum is obtained 
by solving the linearized equations of motion of magnetization. 
The results show that 
the sensitivity of the AHE magnetic sensor can be strongly enhanced 
by tuning the $K_{2}$ of the IP state 
close to the boundary 
between the IP and conically magnetized states 
without degrading the detectivity.

\begin{figure}
\includegraphics[width=0.9\columnwidth]{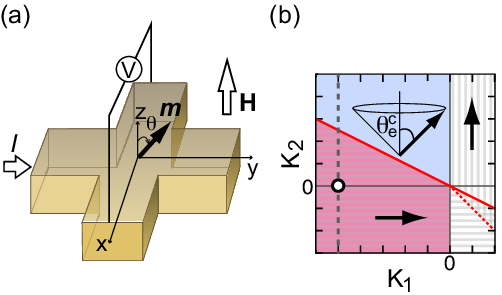}
\caption{
(a) A schematic illustration of an AHE magnetic sensor. 
The direct current, 
$I$, 
and an external field, 
$H$, 
are applied along the $y$-axis and the $z$-axis, 
respectively. 
The Hall voltage is measured along the $x$-axis. 
$\bm{m}$ represents the magnetization unit vector. 
The polar angle, 
$\theta$, 
of $\bm{m}$ is measured from the $z$-axis.
(b) A phase diagram of the equilibrium magnetization 
at $H=0$  
as a function of $K_{1}$ and $K_{2}$. 
The red solid line indicates the boundary 
between the cone and IP states, 
i.e., 
$K_{2}=-K_{1}/2$. 
The cone and IP states appear in the area 
above and below the red solid line, 
the blue- and red-shaded areas, 
respectively. 
Between the red solid and red dashed lines, 
the OP and IP states co-exist. 
Along the dashed vertical line, 
one can change the magnetic state 
by varying $K_{2}$ 
with the same value of $K_{1}$. 
The open circle indicates the IP state with $K_{2}=0$.
\label{fig1}
}
\end{figure}

Figure \ref{fig1}(a) shows a schematic illustration of an AHE magnetic sensor.  
A crossbar-shaped magnetic thin film is laid 
on the $x$--$y$ plane, 
and the $z$-axis is set along the OP direction. 
A direct current is applied along the $y$-axis. 
The Hall voltage generated by the AHE 
is measured along the $x$-axis. 
The external magnetic field, 
$\bm{H}$, 
is applied along the $z$-axis 
as a signal field. 
The polar angle $\theta$ 
and the azimuthal angle $\phi$ 
(not shown in the figure) 
of the unit vector of magnetization 
$\bm{m}$=($\sin\theta\cos\phi$,$\,\sin\theta\sin\phi$,$\,\cos\theta$) 
are measured from the $z$-axis and $x$-axis, 
respectively.

The cross-region where the anomalous Hall voltage is generated 
is assumed to be uniformly magnetized 
because of strong ferromagnetic exchange coupling.  
Representing the direction of magnetization by a unit vector $\bm{m}$, 
the magnetic free-energy density in the cross-region is given by
\begin{align}
E_{H}=K_{1}(1-m_{z}^{2}) + K_{2}(1-m_{z}^{2})^{2} -\mu_{0} M_{s} H m_{z},
\label{eqEnergy}
\end{align}
where $K_{1}$ and $K_{2}$ are the first- and second-order 
uniaxial magnetic anisotropy constants, 
respectively. 
The shape anisotropies in the $x$ and $y$ directions are neglected 
for simplicity. 
The shape anisotropy in the OP direction is included in $K_{1}$,
which can be tuned by varying the film thickness. 
$\mu_{0}$ is the permeability of vacuum, 
$M_{s}$ is the saturation magnetization, 
and $H$ is the external magnetic field to be detected, 
i.e., 
the signal field.

Figure \ref{fig1}(b) shows the phase diagram of the equilibrium magnetization state
in the $K_{1}$--$K_{2}$ plane 
at $H=0$. 
The red solid line shows $K_{2} = - K_{1}/2$, 
indicating the boundary 
between the cone and IP states, 
which is determined by minimizing the magnetic free-energy density with respect to $m_z$.
\cite{casimir_rapport_1959}
The red dashed line indicates the boundary of the region 
where the IP and OP magnetized states co-exist. 
The cone state appears 
when 
$K_1<0$ and
$K_{2} > - K_{1}/2$ are satisfied, 
as shown by the blue shaded area. 
The IP state appears 
when $K_{2}<-K_{1}/2$ is satisfied, 
as shown by the red shaded area.

The equilibrium direction of the cone state 
is given by 
$\bm{m}_{c}=(0, \,\sin\theta_{c}, \,\cos\theta_{c})$, 
where $\theta_{c}=\arccos \left(m_{c,z}\right)$
and
\begin{align}
m_{c,z}= \sqrt{\frac{K_{1}+2K_{2}}{2K_{2}}}.
\label{mzc}
\end{align}
Since the shape anisotropies in the $x$ and $y$ directions are neglected, 
the equilibrium states are independent of $\phi$. 
Hereafter, 
we assume that $\bm{m}$ is in the $y-z$ plane, 
i.e., 
$\phi=\pi/2$, 
for simplicity. 
The equilibrium IP state is assumed to be 
$\bm{m}_{\rm e}^{\rm IP}=(0, \,1, \,0)$.

The Hall resistance of an AHE sensor is given by
\begin{align}
    R=R_{0}+ R_{1}m_{z},
    \label{eqR}
\end{align}
where $R_{0}$ is the Hall resistance independent of magnetization, 
and $R_{1}$ is a coefficient representing the contribution 
from the AHE. 
The following parameters are assumed for numerical calculations. 
The saturation magnetization is $M_{s}=1$ MA/m, 
the first-order uniaxial magnetic anisotropy constant is $K_{1}=-100$ kJ/m$^3$,
the Gilbert damping constant is $\alpha = 0.05$, 
which are chosen to represent typical values for 3d-element 
based ferromagnetic metal.
The sensing area is assumed to be a square of 1 $\mu$m length and 2 nm thickness,
resulting in the effective volume as $\Omega_{\rm v}=2\times10^6$ nm$^3$.
The current density is assumed to be the order of $10^{10}$ A/m$^2$,\cite{gu_size_2024}
which yields the direct current as $I=20$ $\mu$A. 
The magnetization-dependent component of resistance is assumed to be independent of $K_{2}$,
and $R_{1}=0.9$ $\Omega$,
which is estimated by the Hall angle of Fe. \cite{nakatani_r2_06_02_perspective_2024}
The temperature is $T=300$ K.

\begin{figure}
\includegraphics[width=0.9\columnwidth]{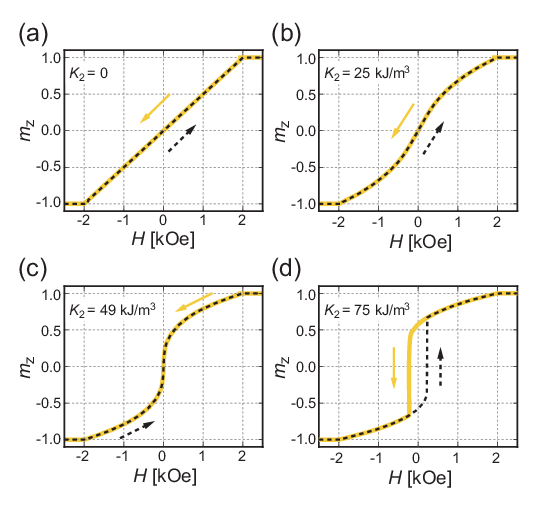}
\caption{
(a) The magnetization curve for $K_{2} = 0$.
The \textcolor{red}{yellow} solid curve illustrates the change in $m_z$ 
with decreasing $H$. 
The \textcolor{red}{black dashed} curve illustrates the change in $m_z$ 
with increasing $H$.
(b) The same plot for $K_{2}$ = 25 kJ/m$^3$.
(c) The same plot for $K_{2}$ = 49 kJ/m$^3$.
(d) The same plot for $K_{2}$ = 75 kJ/m$^3$.
For all panels, $K_{1}=-100$ kJ/m$^3$.
\label{fig2}
}
\end{figure}

Figures \ref{fig2}(a), 
\ref{fig2}(b), 
\ref{fig2}(c), 
and \ref{fig2}(d) 
show the magnetization curves 
for $K_{2} = 0$, 
25, 
49, 
and 75 kJ/m$^{3}$, 
respectively. 
The equilibrium states are obtained 
by calculating the minima of Eq. \eqref{eqEnergy} 
with increasing or decreasing $H$. 
The \textcolor{red}{yellow} solid and \textcolor{red}{black dashed} curves represent $m_{z}$ 
with decreasing and increasing $H$, 
respectively. 
At $H = 0$, 
the system is IP magnetized at $K_{2} = 0$, 
25, 
49 kJ/m$^{3}$ 
and is conically magnetized at $K_{2} = 75$ kJ/m$^{3}$. 
In panels (a), 
(b), 
and (c), 
there is no hysteresis, 
and the slant at $H = 0$ 
increases with increasing $K_{2}$, 
which implies $m_{z}$ becomes more sensitive to $H$ 
with increasing $K_{2}$. 
In panel (d), 
the magnetization curve shows the hysteresis around $H=0$, 
where the cone states with positive and negative $m_{z}$ 
are energetically stable. 
The AHE magnetic sensor with a conically magnetized state 
suffers from a random telegraph noise 
originating from transitions 
between these two states.

\begin{figure}
\includegraphics[width=0.9\columnwidth]{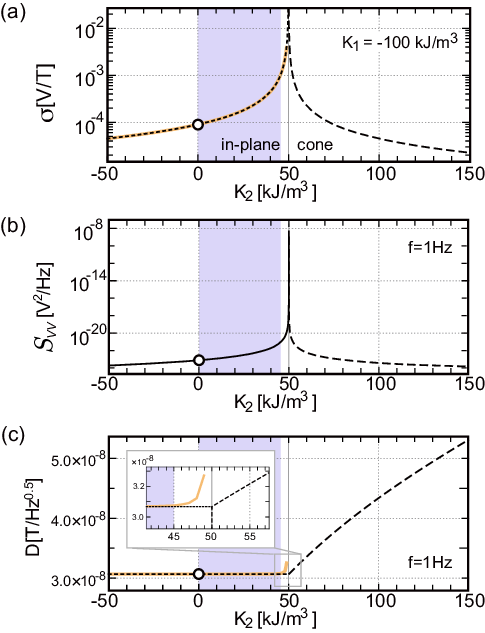}
\caption{
(a) The sensitivity as a function of $K_{2}$
with $K_{1}=-100$ kJ/m$^3$.
The vertical thin line along $K_{2}=50$ denotes the boundary 
between the IP and cone states.
The black dotted and dashed curves represent Eqs. \eqref{eqSigmai}
and  \eqref{eqSigmac},
respectively.
The yellow thick curve represents the sensitivity at $T=300$ K.
The white circle on the solid line at $K_{2}=0$ is intended to guide the eye.
The purple shaded area indicates the optimal range of $K_{2}$ for highly sensitive AHE sensors.
(b) The power spectrum of the voltage noise at $f=1$ Hz.
(c) The detectivity at $f=1$. 
The inset shows a magnified view in the vicinity of the boundary.
\label{fig3}
}
\end{figure}

For the IP state 
satisfying $K_{1}<0$ 
and $K_{2}<-K_{1}/2$, 
the equilibrium value of $m_{z}$ at small $H$ is obtained as
\begin{align}
    \label{eqm0z}
    m_{z} = \frac{\mu_{0}M_{s}}{2(K_{1}+2K_{2})}H + O(H^{2}).
\end{align}
The sensitivity is defined by the derivative of the resistance 
in terms of $H$ as
\begin{align}
    \sigma
    =
    \left |
    \frac{d\langle I R\rangle}{d(\mu_0 H)}
    \right |,
    \label{defsigma}
\end{align}
whose unit is V/T.
One can change the unit of the sensitivity to $\Omega$/T 
by dividing it by the current, which is often used in many literatures.
The sensitivity of the IP state is obtained as
\begin{align}
    \sigma_{\rm IP}=\frac{I R_{1} M_{s}}{|2(K_{1}+2K_{2})|},
    \label{eqSigmai}
\end{align}
which grows 
as $K_{2}$ approaches the boundary
between the IP and the cone states 
and diverges in the limit of $K_{2}\to -K_{1}/2$, 
as shown by the black dotted curve
in Fig. \ref{fig3}(a).
Note that Eq. \eqref{eqSigmai} crresponds to $\sigma_{\rm IP}$ at $T=0$ K.

The power spectrum of the voltage noise,
$S_{VV}$, 
at frequency $f$ can be calculated 
following Refs. \onlinecite{Smith2001,Safonov2002,imamura_r4_1_2024}. 
Introducing the angular frequency $\omega=2\pi f$, 
it is expressed as
\begin{align}
    S_{VV}(\omega)
    =
    4 I^2 \int_{0}^{\infty}
    \langle
    R(t)R(0)
    \rangle
    \cos(\omega t) dt,
    \label{eqSVV}
\end{align}
where $\langle\ \rangle$ represents the statistical average.
Substituting Eq. \eqref{eqR} 
into Eq. \eqref{eqSVV}, 
we obtain
\begin{align}
    S_{VV}(\omega)
    =
    (I R_{1})^2
    S_{m_{z}m_{z}}(\omega),
\end{align}
where $S_{m_{z}m_{z}}(\omega)$ is the power spectrum of $m_{z}$,
defined as
\begin{align}
    S_{m_{z}m_{z}}(\omega)
    =
    4 \int_{0}^{\infty}
    \langle
    m_{z}(t)m_{z}(0)
    \rangle
    \cos(\omega t) dt.
    \label{eqSmm}
\end{align}

The dynamics of $m_{z}(t)$ are obtained 
by solving the following equations of motion,
called the Landau-Lifshitz-Gilbert equation,
\cite{Landau8,Gilbert2004} 
\begin{align}
    \dot{\bm{m}}(t)
    =
    -\gamma\,\bm{m}(t)\times\left\{\bm{H}_{\mathrm{eff}}(t) + \bm{r}(t)\right\}
    +\alpha\, \bm{m}(t)\times\dot{\bm{m}}(t),
    \label{eqLLG}
\end{align}
where $\dot{\bm{m}}(t)$ is the time derivative of $\bm{m}(t)$,
$\gamma$ is the gyromagnetic ratio,
and $\alpha$ is the Gilbert damping constant.
The effective magnetic field, 
$\bm{H}_{\rm eff}$, 
is given by
\begin{align}
    \bm{H}_{\rm eff}
    =\left( \bar{H}_1 m_z -H_2 m_z^3 \right) \bm{e}_z,
\end{align}
where $\bar{H}_1=H_1 + H_2$,
$H_1=2 K_{1} /(\mu_0 M_{s})$,
$H_2=4 K_{2} /(\mu_0 M_{s})$,
and $\bm{e}_{z}$ is the unit vector 
pointing in the positive $z$ direction.
Note that
$\bar{H_1}<0$ under the condition of the IP state,
$K_{1} + 2 K_{2} < 0$.
The thermal agitation field, 
$\bm{r}$, 
is determined 
by the fluctuation-dissipation theorem
\cite{callen_43_irreversibility_1951,callen_44_on_1952,callen_45_statistical_1952,greene_46_on_1952,brown_42_thermal_1963}
and satisfies the following relations: 
$\langle r_{i}(t)\rangle = 0$ 
and 
\begin{align}
     & \langle r_{i}(t)r_{j}(t')\rangle = \xi\delta_{ij}\delta(t-t'),
    \label{eqRcorr}
\end{align}
where
the indices $i$ and $j$ denote the $x$, $y$, and $z$ components 
of the thermal agitation field,
$\delta_{ij}$ represents the Kronecker delta,
and $\delta(t-t')$ represents Dirac's delta function.
The coefficient $\xi$ is given by
\begin{align}
    \xi=\frac{2\alpha\,k_{B}T}{\gamma\,\mu_0\, M_{s}\,\Omega_{\rm v}},
    \label{eqXi}
\end{align}
where 
$k_{B}$ is the Boltzmann constant,
$T$ is the temperature,
and $\Omega_{\rm v}$ is the effective volume of the sensing region.

Assuming $|m_x|\ll 1$, 
$|m_{z}| \ll1$, 
and $|m_{y}| \simeq 1$,
the following linearized equations of motion are obtained:
\begin{align}
    \dot{m}_x(t)   & =-\gamma\bar{H}_1 m_z(t) -\gamma r_{z}(t)        \\
    \dot{m}_{y}(t) & =0                                               \\
    \dot{m}_{z}(t) & =\gamma r_{x}(t)  +\alpha\gamma\bar{H}_1 m_z(t).
\end{align}
These linearized equations of motion can be solved 
by using the Fourier transform of $m_{i}(t)$, 
$(i=x,y,z)$ 
defined as $m_{i}(\omega)=\int_{-\infty}^{\infty}m_{i}(t)\exp(-\omega t)dt$. 
After some algebra, 
we obtain
\begin{align}
     & m_{z}(\omega)
    = \frac{\gamma \, r_x(\omega)}{i\omega - \alpha\gamma \bar{H}_1}.
\end{align}
Substituting the inverse Fourier transformation of 
$m_{z}(\omega)$ 
into Eq. \eqref{eqSmm}, 
and noting that 
the correlation of the thermal agitation field 
in Fourier space is given by \cite{brown_42_thermal_1963}
\begin{align}
    \langle
    r_i(\omega) r_j(\omega')
    \rangle
    =
    2\pi \xi \delta_{ij}\delta(\omega + \omega'),
\end{align}
the power spectrum of voltage noise for the IP state is obtained as
\begin{align}
    S_{VV}^{\rm IP}(\omega)
     & =
    \frac{2I^2 R_{1}^2 \gamma^2 \xi}{\omega^2 +F^2  },
    \label{eqSi}
\end{align}
where 
$F = \alpha \gamma \bar{H}_1$.
As shown in Fig. \ref{fig3}(b), 
$S_{VV}^{\rm IP}(\omega)$ increases 
with increasing $K_{2}$ 
and takes the maximum value of 
$\left(2I^2 R_{1}^2 \gamma^2 \xi\right)/\omega^2$ 
at $K_{2}=-K_{1}/2$.

Figure \ref{fig3}(c) shows the detectivity, 
defined as
\begin{align}
    D
    =\frac{\sqrt{S_{VV}(\omega)}}{\sigma},
    \label{Ddef}
\end{align}
which represents the noise 
converted into the magnetic fields. 
A smaller $D$ means less noise in a sensor.
From Eqs. \eqref{eqSigmai} and \eqref{eqSi},
the detectivity of the IP state is obtained as
\begin{align}
    D_{\rm IP}=\frac{\mu_0\sqrt{2\xi}}{\alpha}\frac{1}{\sqrt{1 + (\omega/F)^2}  }.
\end{align}
It should be noted that 
$D_{\rm IP}$ is almost independent of $\omega$ 
for a low-frequency region, 
i.e., 
$\omega \ll {F}$,
as shown by the black dotted line in Fig. \ref{fig3}(c),
and equals zero at the boundary of $K_{2}$ 
with $\omega\ne0$
(see the inset). 
For the parameters, 
we assumed $K_{2}=49$ kJ/m$^3$
and $F$ is of the order of $10^7$ rad/s.

For the cone state, 
the sensitivity, 
power spectrum of voltage noise,
and detectivity can be calculated 
in the same manner as for the IP state.
Neglecting the transition 
between the cone states with positive and negative $m_{z}$, 
a straightforward calculation yields
\begin{align}
    \sigma_{\rm c}=\frac{I R_{1} M_{s}}{4(K_{1}+2K_{2})},
    \label{eqSigmac}
\end{align}
\begin{align}
    S_{VV}^{\rm c}(\omega)
    = 2 I^2 R_{1}^2 \gamma^2 \xi \left(\frac{-K_{1}}{2K_{2}}   \right)
    \frac{1}{\omega^2 + \hat{U}^2},
    \label{eqSc}
\end{align}
and
\begin{align}
    D_{\rm c}
     & =
    \frac{\mu_0\sqrt{2\xi}}{\alpha}
    \sqrt{\frac{2K_{2}}{-K_{1}}}
    \frac{1}{{\sqrt{1 + (\omega/\hat{U})^2}}},
    \label{eqDc}
\end{align}
where 
$\hat{U}=2\alpha\gamma H_2(\sin\theta_{c}\cos\theta_{c})^2$.

The results of the cone state satisfying $K_{1}<0$ 
and $K_{2}>-K_{1}/2$ 
are shown by the dashed curves 
in Figs. \ref{fig3}(a)--\ref{fig3}(c). 
Similar to the IP state, 
the sensitivity and the power spectrum of voltage noise are strongly enhanced 
in the vicinity of the boundary 
between the IP and cone states, 
as shown in Figs. \ref{fig3}(a) and \ref{fig3}(b). 
In contrast to the IP state, 
the detectivity increases with increasing $K_{2}$, 
as shown in Fig. \ref{fig3}(c). 

Let us discuss the effect of temperature on the sensitivity and detectivity.
The working temperature should affect the sensitivity through the Boltzmann distribution in the statistical average. 
The sensitivity at $T=300$ K of IP state is shown by the yellow solid curve in Fig. \ref{fig3}(a).
The line is almost the same as $\sigma_{\rm IP}$ 
except in the very vicinity of the boundary point of $K_2=-K_1/2$,
because the size of the sensing area is assumed to be large 
enough to inhibit the effects of temperature.
The divergence of the sensitivity at the boundary point is suppressed at finite temperature, 
which results in the small enhancement of the detectivity in the very vicinity of the boundary point,
as shown by the yellow curve in the inset of Fig. \ref{fig3}(c).

From the results shown in Figs. \ref{fig2} and \ref{fig3}, 
it can be concluded that 
the optimal range of $K_{2}$ for highly sensitive AHE sensors
is the purple shaded region in Fig. \ref{fig3}.

It would be useful to comment briefly on the possible materials having the IP state with $K_2>0$ and cone state. It is known that the cone state can appear in the CoFeB/MgO system\cite{timopheev_inhomogeneous_2017, bultynck_instant-spin_2018} and Pt/Co multilayer system\cite{stillrich_magnetic_2009} by controlling the thickness of magnetic layer such as CoFeB and Co. These materials are also found to be in the IP state with $K_2>0$.
The ion irradiation is alternative approach to control the magnetic states.
Toxerira {\it et al.}, have demonstrated that different ion irradiation intensities can produce distinct magnetic states, namely IP, cone and OP states in the same multilayer stacking structure comprising a MgO/Fe$_{72}$Co$_{8}$B$_{20}$/X(0.2 nm)/Fe$_{72}$Co$_{8}$B$_{20}$/MgO layer stack, where X stands for an ultrathin Ta or W spacer\cite{teixeira_ion_2018}.

One of the major advantages of the method of tuning $K_2$ is that the conventional spintronics materials such as CoFeB can be used, and $K_2$ in these materials can be easily controlled by changing the thickness of thin film or using ion irradiation. Most materials with large anomalous Hall resistivity consist of heavy noble metal atoms such as Pt and Ir, which are expensive. In addition, little is known about the method to control $K_2$ in these materials

In summary, 
we have theoretically analyzed the effect of the second-order anisotropy constant 
on the sensitivity, 
power spectrum of voltage noise, 
and detectivity of AHE magnetic sensors. 
We have found that 
the sensitivity of an IP-magnetized AHE magnetic sensor can be strongly enhanced 
without degrading the detectivity 
by tuning the value of $K_{2}$ 
close to the boundary 
between the IP and the cone states. 
The results of this study provide a foundation 
for the development of highly sensitive AHE sensors 
that do not require materials 
exhibiting large anomalous Hall resistances.

\section*{ACKNOWLEDGMENTS}
We would like to thank T. Nakatani for valuable discussions.
\section*{AUTHOR DECLARATIONS}
\subsection*{Conflict of interest}
The authors have no conflicts to disclose.
\subsection*{DATA AVAILABILITY STATEMENT}
The data that support the findings of this study are available from the corresponding author upon reasonable request.
%
  
\end{document}